# Proximity-induced ferromagnetism and chemical reactivity in few layers VSe$_2$ heterostructures

G. Vinai,[1] C. Bigi,[1,2] A. Rajan,[3] M. D.Watson,[3] T.-L. Lee,[4] F. Mazzola,[3] S. Modesti,[1,5] S. Barua,[6,7] M. Ciomaga Hatnean,[6] G. Balakrishnan,[6] P. D. C. King,[3] P. Torelli,[1] G. Rossi,[1,2] G. Panaccione[1]

[1] Istituto Officina dei Materiali (IOM)-CNR, Laboratorio TASC, Area Science Park, S.S. 14 km 163.5, Trieste I-34149, Italy
[2] Department of Physics, Università degli Studi di Milano, Via Celoria 16, I-20133 Milano, Italy
[3] SUPA, School of Physics and Astronomy, University of St. Andrews, St. Andrews KY16 9SS, United Kingdom
[4] Diamond Light Source Ltd., Harwell Campus, Didcot OX11 0DE, United Kingdom
[5] Department of Physics, Università di Trieste, via Valerio 2, I-34127 Trieste, Italy
[6] Department of Physics, University of Warwick, Coventry CV4 7AL, United Kingdom
[7] Department of Physics, Birla Institute of Technology, Mesra 835215, Ranchi, India

## Abstract

Among Transition-Metal Dichalcogenides, mono and few-layers thick VSe$_2$ has gained much recent attention following claims of intrinsic room-temperature ferromagnetism in this system, which have nonetheless proved controversial. Here, we address the magnetic and chemical properties of Fe/VSe$_2$ heterostructure by combining element sensitive absorption spectroscopy and photoemission spectroscopy. Our x-ray magnetic circular dichroism results confirm recent findings that both native mono/few-layer and bulk VSe$_2$ do not show any signature of an intrinsic ferromagnetic ordering. Nonetheless, we find that ferromagnetism can be induced, even at room temperature, after coupling with a Fe thin film layer, with antiparallel alignment of the moment on the V with respect to Fe. We further consider the chemical reactivity at the Fe/VSe$_2$ interface and its relation with interfacial magnetic coupling.

## 1. Introduction

2D Transition Metal Dichalcogenides (2D-TMDs) have been recently attracting increasing attention due to their unique physical properties when passing from three-dimensional crystals to single or few layers [1–5], with applications ranging in electrocatalysis [6,7], optoelectronics [4], batteries [8,9], piezoelectricity [10] and memory devices [11].

Among dimensionality effects observed in TMDs, metallic VSe$_2$ is a paradigmatic case. While in its bulk form it displays the onset of charge density waves (CDW) at 110 K [12–14], in the 2D limit a CDW with enhanced T$_C$ and coupling strength is observed [15–17] (sometimes even considered a metal-insulator transition [15]), and with a different pattern of atomic displacements to the bulk [18,19]. Moreover, its dimensionality-dependent magnetic properties are still under debate. First-principles calculations have predicted that monolayer (ML) VSe$_2$ might be a 2D itinerant-type ferromagnet, with a magnetic moment per unit cell of about 0.69 $\mu_B$ [1,16,20–23]. Although magnetometry measurements reported a strong ferromagnetic response at room temperature [20,24,25], element sensitive spectroscopic results showed no magnetic signal at V $L_{2,3}$ edges, down to cryogenic temperatures [26,27]. Only very recently, a dichroic signal has been reported in the case of chemically exfoliated ML VSe$_2$, which becomes more pronounced after surface passivation.[28] Thus, while the most recent studies generally find that quasi-freestanding monolayer VSe$_2$ is not intrinsically ferromagnetic [18], it remains an interesting open question if ferromagnetism can be coupled into the system, for instance by proximity to a magnetic overlayer.

In this work, we explore the use of such magnetic proximity effects where the magnetic coupling with a ferromagnetic overlayer (Fe in present case) is used as spectroscopic fingerprint, an approach already used in other 2D systems [17,29–31]. Specifically, we aim at investigating both the intrinsic and the induced magnetism of VSe$_2$, including its layer dependent behaviour, before (i.e. in the pristine case) and after depositing a thin Fe layer on top of it. In our study, based on chemical sensitive techniques such as X-ray absorption spectroscopy (XAS), X-ray magnetic circular dichroism (XMCD), and photoemission spectroscopy (PES), we focus on two aspects: (i) the magnetic state of VSe$_2$, both as bulk and MLs, and (ii) the chemical stability and/or chemical modifications occurring at the Fe/VSe$_2$ interface. While XMCD measured at V $L_{2,3}$ edges on pristine 3D and 2D VSe$_2$ do not display magnetic signal at room temperature [26,27], a clear ferromagnetic signal at room temperature is observed when a Fe overlayer is deposited, with an antiparallel coupling between V and Fe. We also identify a tendency of Se to migrate towards the surface, leading to a (at least partial) metallization of V and therefore to a Fe/V antiferromagnetic coupling. Our element sensitive characterizations of the Fe/2D-VSe$_2$ heterostructure open the way to further studies on proximity effects on 2D-TMDs, with the aim of reducing the chemical reactivity at the interface and maximizing the proximity-induced magnetic coupling.

The paper is organized as follows. Section II briefly describes the growth techniques of the MLs and bulk VSe$_2$, the decapping procedure and the Fe deposition conditions, together with the experimental setup of the measurements. Section III will focus on the decapped, pristine VSe$_2$ samples, whereas Section IV will concern the study on both the proximity effect and chemical reactivity of the Fe/VSe$_2$ heterostructure. Finally, Section V will summarize the results and draw the main conclusions.

## 2. Methods

2D-VSe$_2$ films were grown on graphene (Gr) / SiC substrates using molecular-beam epitaxy (MBE) deposition. Details on the growth technique can be found in ref. [26]. The number of MLs of the different samples was estimated by calibrating the deposition rate. Samples were then capped with a protective Se layer of few nanometers thickness after growth to allow transferring them in air. Single crystals of VSe2 were produced by the Chemical Vapour transport technique [32]. The VSe$_2$ bulk sample was cleaved in ultra high vacuum conditions (base pressure $\sim 1\times 10^{-10}$ mbar).

XAS, XMCD and PES measurements were performed at APE-HE beamline at Elettra synchrotron [33]. XAS and XMCD measurements were taken in total electron yield (TEY) mode, normalizing the intensity of the sample current to the incident photon flux current at each energy value. Absorption spectra were taken in circular polarization, with an incident angle of 45°. The XMCD measurements were performed under remanence conditions: i.e. at each energy point of the spectra, alternating magnetic field pulses of ± 300 Oe were applied in the sample plane (exceeding the field strength at which the magnetisation saturates), and then the signal measured in zero applied field in both cases. Dichroic signal intensities were corrected by taking into account both the 75% degree of circular polarization of the incident light and the 45° between the sample magnetization and the photon angular momentum. The spectra were taken both at room temperature and at 100 K. Element sensitive hysteresis loops at the V and Fe edges were taken by selecting the $L_3$ edge and pre-edge absorption energies of either V or Fe with both phonon helicities and scanning the magnitude of the magnetic field in the range ±100 Oe in the sample plane. PES measurements were recorded with an Omicron EA125 hemispherical electron energy analyser, with the sample at 45° with respect to the impinging linearly polarized light and normal to the surface.

Further experiments were performed at I09 beamline at Diamond Light Source (UK), including LEED, PES and ARPES measurements at soft-X ray energies. PES and ARPES measurements were recorded with a VG Scienta EW4000 analyser, at 75 K in the latter case. The endstation is designed with an angle between the incident beam and the analyser axis of 87°. Angular dependent PES measurements were taken with an incident angle of 39°, with a ± ~20° range angular dependence, which gives an emission angle (i.e. the angle between the emitted electron and the sample surface normal) between 70° (more grazing, i.e. more surface sensitive) and 30° (more normal, i.e. more bulk sensitive).

After the decapping of the samples, in both experiments the Fe deposition on ML $VSe_2$ thin films and bulk samples was done via MBE in a preparation chamber connected to the end stations chambers, at a base pressure of $2 \times 10^{-10}$ mbar, with a deposition rate of 0.65 Å/min. All Fe depositions were done at room temperature.

## 3. Pristine $VSe_2$

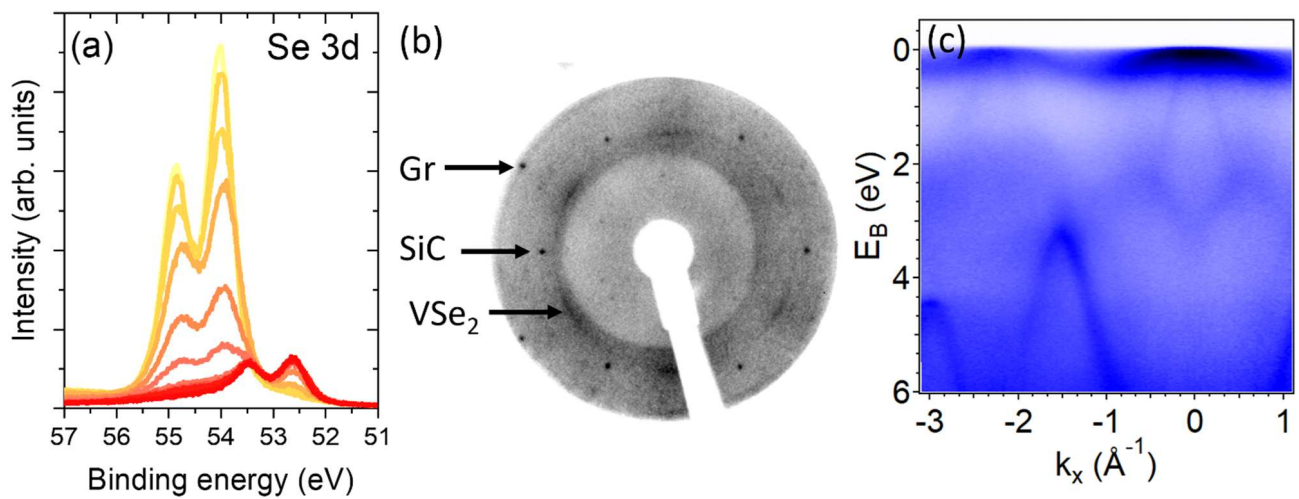

**Figure 1** – (a) PES measurements at 900 eV of Se 3d edges of a capped 1 ML $VSe_2$ sample during the decapping process. The sample is heated up during the measurements (yellow to red for increasing temperature). (b) LEED pattern after decapping taken at room temperature at 108 eV. (c) ARPES measurement taken at 75 K at 110 eV.

The decapping of a protective Se layer deposited atop the MBE-grown samples after growth was done by slowly heating up the sample up to ~450 K in situ and monitoring via PES the evolution of V 2p and Se 3d edges during the decapping process, with a impinging photon energy of 900 eV. **Figure 1** shows an example of the evolution of the Se 3d edges during the decapping. Its initial state (in yellow) corresponds to an amorphous $Se^0$ state, with its peaks at 54.9 and 54 eV. While heating, the peaks gradually shift, since reaching the $Se^{2-}$ state (in red), with peaks at 53.5 and 52.6 eV for Se $3d_{5/2}$ and $3d_{3/2}$ respectively, for a shift of 1.4 eV between the two states, in good accordance with previously reported PES studies on $VSe_2/Se^0$ decapping measurements [34]. At the same time, V 2p peaks increase in intensity as long as the temperature increases (Figure S1 in the Supplementary material). The initial peaks present shoulders, probably due to contamination coming from the air, which disappear once the capping is fully removed. The final V 2p edges have $2p_{3/2}$ and $2p_{1/2}$ peaks at 513 and 520.5 eV, i.e. with a spin-orbit coupling of 7.5 eV [27].

The quality of the complete decapping of the surface and of the 2D-VSe$_2$ was verified via LEED and ARPES measurements (Figures 1b and 1c). The LEED pattern on clean 2D-VSe$_2$ (Figure 1c), measured at room temperature at 108 eV, shows (i) the sharp spots coming from both the Gr underlayer and the SiC substrate, and (ii) the elongated Bragg spots of the 2D-VSe$_2$ layer, due to the distributed rotational domains of the ML, consistent with recent observations in similar systems [26]. The ARPES measurements, taken at 75 K at 110 eV (Figure 1d) along K-Γ-M axis, shows the V 3d band localized close to the Fermi edge and the Se 4p bands dispersing downward, centred on $k_x$=0 [26,27,35].

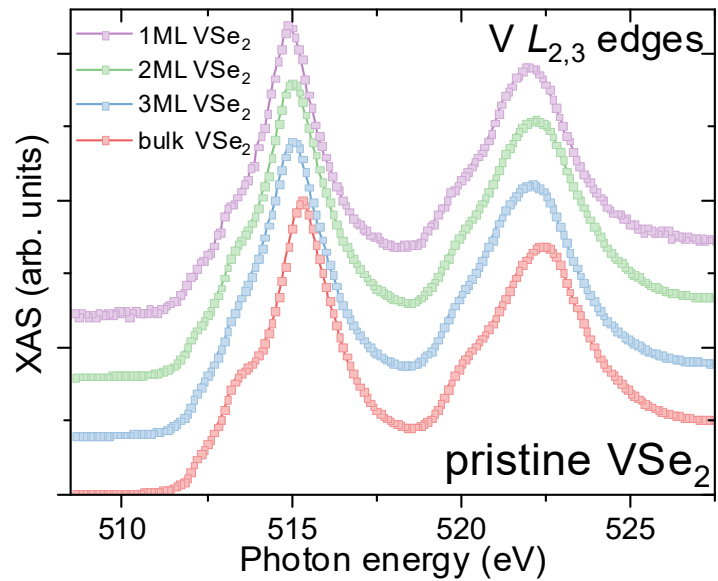

**Figure 2** – XAS measurements at V $L_{2,3}$ edges on VSe$_2$ MLs after the decapping procedure and VSe$_2$ bulk after cleaving. All spectra were taken at room temperature.

**Figure 2** shows the XAS spectra at V $L_{2,3}$ edges taken on both bulk and MLs VSe$_2$ films after the removal of capping Se layer. Both 3D and 2D samples present a 4+ state, with $L_{2,3}$ peaks of 2D samples with energies shifted to lower values of 0.3 eV with respect to the bulk one. All samples present a shoulder before $L_3$ edge, at 513.3 eV, particularly pronounced in the case of bulk sample. $L_{2,3}$ peaks positions at 514.8 and 522 eV respectively for the ML thin films place them at energy values lower than what reported for VO$_2$ [36–40] and higher than metallic V (reported at 512 eV) [41], whereas the bulk sample has a slightly shifted $L_3$ edge, at 515.3 eV.

Regarding the magnetic behaviour of VSe$_2$ before Fe deposition, no sign of dichroic signal was detected on any of the samples within the instrumental sensitivity limits, confirming what previously reported on similar samples by the same technique [26,27]. In the following, such lack of dichroic signal will be discussed by comparing it with the case of the proximity-induced magnetism in the Fe/VSe$_2$ heterostructure.

## 4. Fe/VSe$_2$ Heterostructure

### 4.1 Proximity induced ferromagnetism

**Figure 3** shows the XAS and XMCD spectra at V $L_{2,3}$ edges taken before (light colours) and after (dark colours) the 2 nm Fe deposition, for both the 1ML case (Figure 3a) and the bulk VSe$_2$ (Figure 3b) sample, together with the XAS and XMCD spectra at Fe $L_{2,3}$ edges of the 1ML case (Figure 3c). All spectra are measured at room temperature.

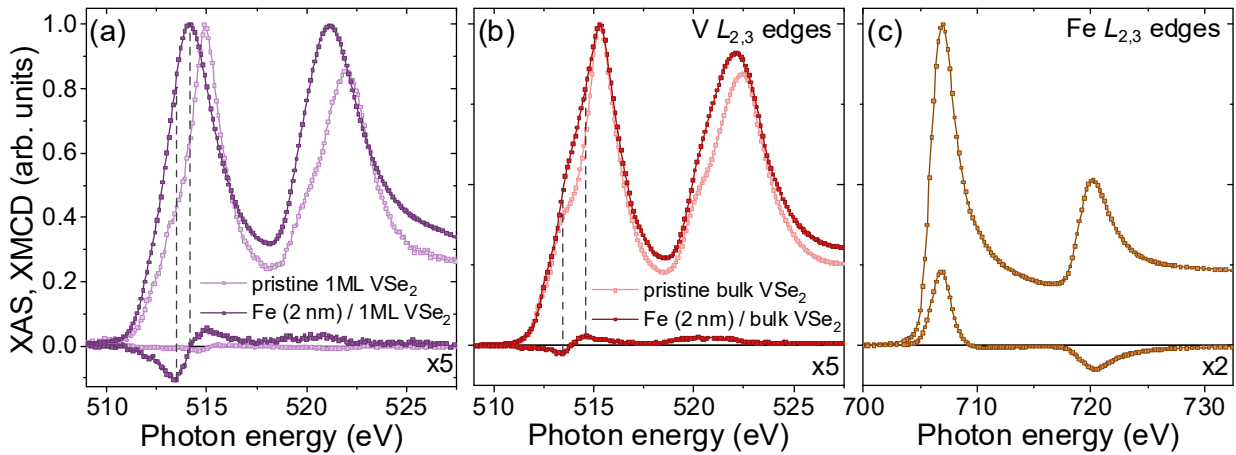

**Figure 3** – (a,b) XAS and XMCD spectra at V $L_{2,3}$ edges on pristine VSe$_2$ (light colours) and Fe (2 nm) / VSe$_2$ (dark colours), for 1ML (a) and bulk (b) samples. XMCD spectra are multiplied by a factor 5. (c) XAS and XMCD at Fe $L_{2,3}$ edges for Fe (2 nm) / 1ML VSe$_2$. All spectra are measured at room temperature.

Firstly, we focus on the comparison of the XAS spectra. In Figure 3a, we can notice that the 1ML VSe$_2$ sample presents a general shift of V $L_{2,3}$ edges towards lower energies after Fe deposition compared to the pristine case, together with a smoothening of the $L_3$ pre-edge features. Similar modifications were observed for all MLs samples. Such shift is an indicator of a possible chemical modification of the ML upon Fe deposition, with a tendency of V to metallize towards a V$^0$ state, as will be further discussed in the following. In the case of the bulk sample (Figure 3b), the position and width of the $L_3$ edge remain unchanged after Fe deposition, whereas the pre-edge is modified, with an increase of its signal intensity.

Regarding the XMCD, Figure 3a shows a comparison between the XMCD signal of 1ML VSe$_2$ sample before and after Fe deposition, both magnified by a factor of 5 with respect to the XAS sum spectra. In the case of the Fe/2D-VSe$_2$ heterostructure, a clear $L_3$ asymmetry peak is measured. The maximum of the asymmetry takes place on the $L_3$ pre-edge at 513.5 eV, with an intensity of 2.3%, while its value goes to zero on the $L_3$ edge (514 eV). In the case of the 2 and 3 MLs samples, their dichroic signal was 1.2% and 1.4% respectively. Together with the XMCD of the heterostructure, Figure 3a shows the dichroic signal of V for the pristine 1ML VSe$_2$. We can observe that no features are present in correspondence to the $L_3$ asymmetry of the heterostructure, while the signal intensity, below 0.3%, is inside the instrumental sensitivity limit [42]. In the case of bulk VSe$_2$, the XMCD features are much less defined than in the 2D cases, with an asymmetry of 0.5%. Interestingly, the photon energy value of the maximum of the asymmetry of the 3D sample (513.5 eV) is the same of the 2D ones. This suggests that for both ML and bulk samples the contribution to the dichroic signal comes from the interface. Whereas in the former case the whole sample is affected by the interfacial coupling because of its 2D nature, in the latter one the TEY probing depth of ~5-7 nm integrates over a thicker volume of the sample, thus the interfacial chemical modifications of V are mostly covered by the unmodified signal coming from below the interface.

Corresponding to the dichroic signal of V, Figure 3c shows the XAS and XMCD spectra at Fe $L_{2,3}$ edges of the same Fe (2 nm) / 1ML VSe$_2$ sample. The $L_3$ dichroic signal intensity at Fe edge is 23%, i.e. the Fe layer thickness was not large enough to have a full Fe magnetization at room temperature [43]. Similar values of Fe dichroic signals were obtained for all samples. V and Fe dichroic signals are opposite in sign, which is indicating an antiparallel coupling between the two in the film plane. A similar antiparallel coupling has been recently observed on Co/2D-VSe$_2$ heterostructures [44].

By using the sum rules [43], it is possible to estimate the total magnetic moment of V in the heterostructure. This operation requires great carefulness because of two aspects: (a) the very close distance between $L_2$ and $L_3$ edges, which increases the error bar of the measured values, and (b) the possible coexistence of metallic V and VSe$_2$. Indeed, the evolution of the XAS V spectra upon Fe deposition opens questions on the chemical stability of the ML, or more generally on the interfacial layer. Indeed, XAS spectra show a tendency of V to metallize towards a $V^0$ state, with the shift of $L_3$ edges towards lower energies. The appearing antiferromagnetic coupling between Fe and V is consistent with what has been observed via XMCD measurements in the case of metallic Fe/V interfaces [45–50]. The shape of the measured XMCD shown in Figure 3 has good resemblance with the dichroic signal reported in the case of Fe/V multilayers [46], which reports a total magnetic moment of V of 0.26 $\mu_B$. In the case of 2D-$VSe_2$, the theoretically predicted total magnetic moment of V is of about 0.69 $\mu_B$ [16]. Because of these considerations, both 4+ and purely metallic V states have been considered. By normalizing the measured Fe dichroic signal for the bulk value of Fe of 2.18 $\mu_B$, we obtain a total magnetic moment of V that has its maximum value for the 1ML case of the order of 0.16 ± 0.08 $\mu_B$, a value closer to the metallic V case than the $VSe_2$ one.

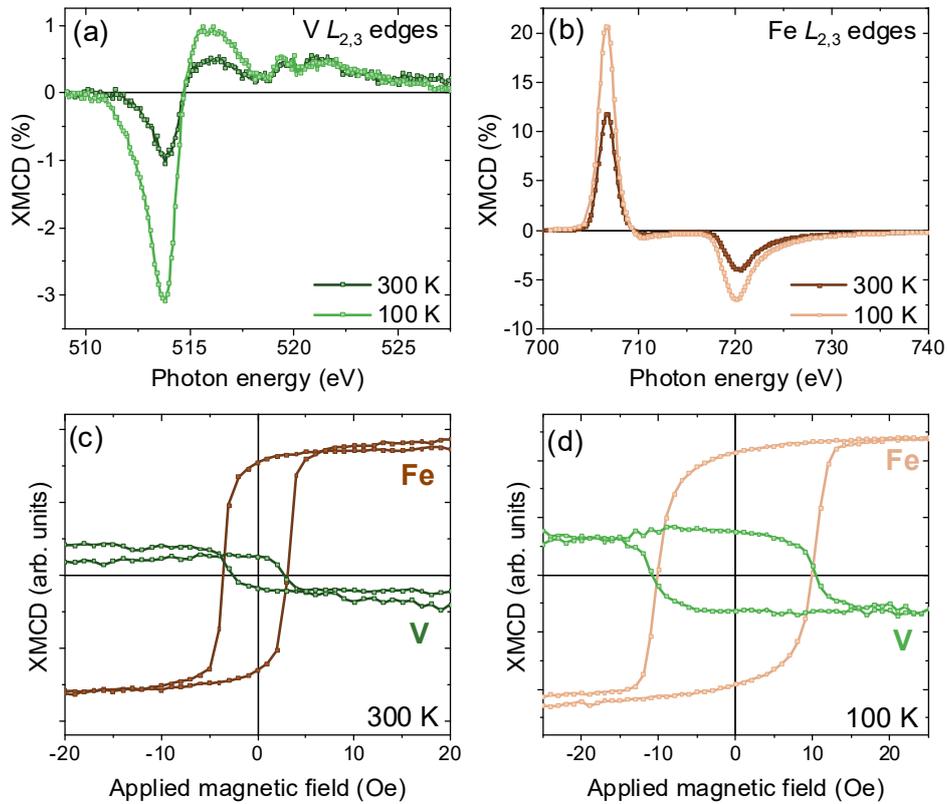

**Figure 4** – (a,b) XMCD spectra at (a) V and (b) Fe $L_{2,3}$ edges on Fe (2nm) / 2ML $VSe_2$ sample at 300 K (dark) and 100 K (light); (c,d) Element sensitive hysteresis loops at Fe and V $L_3$ edges on Fe (2nm) / 2ML $VSe_2$ sample at (c) 300 K and (d) 100 K

To further prove the antiferromagnetic coupling at the interface, we measured element sensitive hysteresis loops on both V and Fe edges, at room temperature and at 100 K. **Figure 4** shows the evolution of the ferromagnetic response of Fe / 2ML $VSe_2$ heterostructure. Figure 4a,b shows the XMCD signals at (a) V and (b) Fe $L_{2,3}$ edges at 300 K and 100 K. Both V and F dichroic signals proportionally increase once cooled down, confirming an interfacial exchange coupling between the two elements. The antiferromagnetic interfacial coupling is moreover confirmed by element sensitive hysteresis loops at Fe and V $L_3$ edges. Figures 4c and 4d show the hysteresis loops of Fe (2nm) / 2ML VSe$_2$ sample at (c) 300 K and (d) 100 K. The signal of the V follows the magnetic response of the Fe layer, with identical coercive field and opposite sign. At room temperature, Fe presents a coercive field of 4 Oe and a ratio between remanence magnetization and saturation magnetization $M_r/M_{sat}$ ratio of ~80%, with V with antiparallel alignment. At 100 K, Fe coercive field reaches 10 Oe, with similar $M_r/M_{sat}$ ratio.

**4.2 Chemical reactivity at Fe/VSe$_2$ interface**

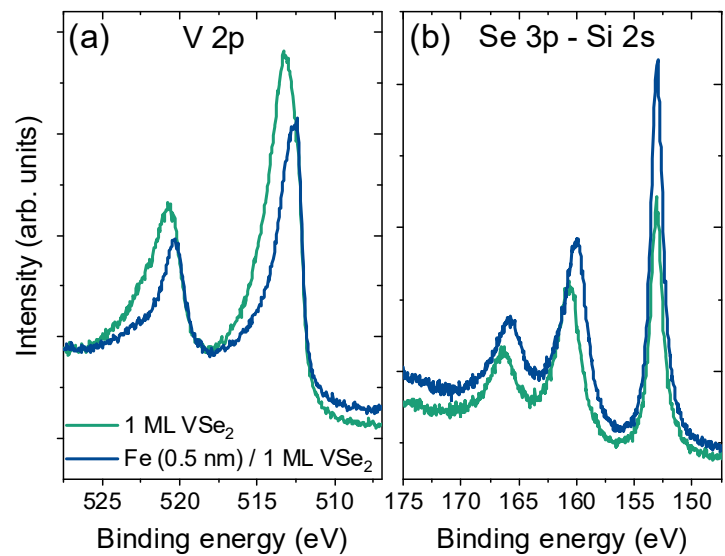

**Figure 5** – PES measurements at 900 eV on 1 ML VSe$_2$ after decapping (green) and after 0.5 nm Fe deposition (blue): (a) V 2p edge, (b) Se 3p and Si 2s edges.

To have an in depth understanding of the interfacial coupling taking place at the Fe/VSe$_2$ interface, surface sensitive spectroscopic characterizations such as PES, ARPES and LEED measurements were taken, comparing a pristine 1ML VSe$_2$ sample and the same sample after 0.5 nm Fe deposition. **Figure 5** shows a comparison of PES spectra, taken at 900 eV before and after 0.5 nm Fe deposition at V 2p (a) and at Se 3p and Si 2s (b) edges. A series of modification at these edges occurring between the two stages suggests how the creation of a Fe/VSe$_2$ interface affects the features of the whole VSe$_2$ ML. Firstly, V 2p peaks shift towards lower binding energies (Figure 5a), with a shift of V 2p$_{3/2}$ of -0.5 eV (from 513.1 eV to 512.6 eV). At the meantime, Se 3p shift towards higher binding energies (Figure 5b), with a shift of Se 3p$_{3/2}$ of +0.5 eV, whereas Si 2s peak, coming from the substrate, remains unmodified. These energy shifts are a signature of an intermixing between VSe$_2$ and Fe after Fe deposition. Regarding Se 3d edge, its 3d$_{3/2}$ and 3d$_{1/2}$ peaks at 54.3 and 53.5 eV measured after decapping, characteristic of VSe$_2$ MLs [34], are overlapping with Fe 3p edge after 0.5 nm Fe deposition. The Se 3d features are therefore not recognizable anymore, and replaced by a broad Fe 3p edge (Figure S2 in the supplementary). The signs of an intermixing between VSe$_2$ and Fe were also confirmed by the LEED and ARPES measurements. After Fe deposition, the LEED pattern measured in the same conditions showed no features. Whereas the spots of the Gr/Si substrate, already weak on the pristine ML VSe$_2$, are hardly detectable since almost out of the probing depth of the measurements, the Bragg spots of VSe$_2$ are not measurable anymore, while no spots due to Fe deposition are detected (Figure S3 in the supplementary). This loss of information implies a loss in details in ARPES features too, which are almost completely covered by the broadly dispersive band of Fe (Figure S3).

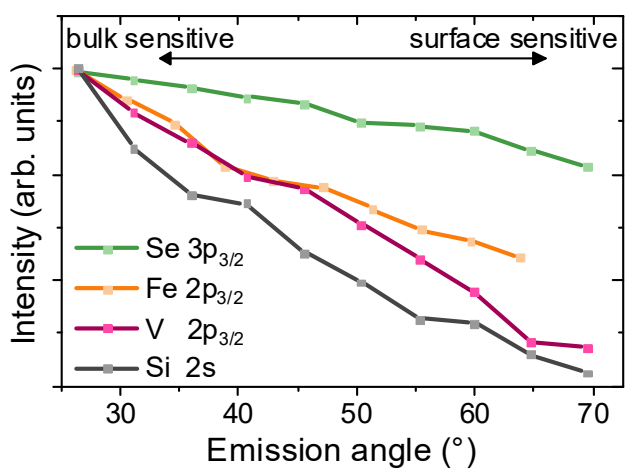

**Figure 6** – Angular dependent evolution of the peak intensities of the PES spectra at Se 3p$_{3/2}$, Fe 2p$_{3/2}$, V 2p$_{3/2}$ and Si 2s peaks on Fe (0.5 nm) / 1 ML VSe$_2$ sample, as a function of the emission angle.

A direct proof of the interfacial intermixing can be seen via an element sensitive depth profile of the Fe (0.5 nm) / 1 ML VSe$_2$ interface, by means of angular dependent PES measurements. **Figure 6** shows the angular dependent evolution of the peak intensities of the PES spectra at Fe 2p$_{3/2}$, Se 3p$_{3/2}$, V 2p$_{3/2}$ and Si 2s edges on Fe (0.5 nm) / 1 ML VSe$_2$ sample, as a function of the emission angle. Measurements were taken by changing the photon energy in order to have a kinetic energy of 250 eV at the main edge for all spectra. This allows having comparable probing depths among the spectra, i.e. correctly comparing the depth profiles of the different elements. Non-uniformities due to the analyser were taken into account by normalizing the curves using a reference background (i.e. a flat photoemission spectrum). Finally, the intensities were corrected by taking into account the photoelectron angular distribution parameters for each element [51]. The intensities were therefore normalized to one at the smallest emission angle, i.e. at the most bulk sensitive measurements.

In case of an element confined at a certain height of the sample stack, the peak intensity is expected to decrease monotonically with the increase of the emission angle, with a larger slope for elements far from the surface.[52] In Figure 6, the Si 2s peak signal (in grey), coming from the substrate, acts as a reference. In the case of Fe $2p_{3/2}$ and V $2p_{3/2}$ peaks, the slope is reduced, with a signal of Fe slightly larger than the one of V at large emission angles. We can therefore consider Fe stably confined on top of the ML and V below the deposited Fe layer. Se $3p_{3/2}$ peak, on the other hand, shows a strongly reduced slope, which indicates the tendency of Se to segregate from the 2D-VSe$_2$ and migrate towards the surface, thus inducing the metallization of V, as shown in the previous paragraph. Se is known to easily form Se-Fe bonding; a similar interfacial chemical reaction has been observed at Fe/ZnSe interface, independently on the ZnSe initial surface termination.[53]

## 5 Conclusions

We have presented an element sensitive characterization of the magnetic and chemical properties of VSe$_2$, from bulk down to few layers, both in the pristine case and in the Fe/VSe$_2$ heterostructure. No intrinsic ferromagnetism, as due to the absence of dichroic signal, is observed on pristine samples, at any thickness. After depositing the Fe overlayer, an antiparallel aligned dichroic signal appears at V and Fe $L_{2,3}$ edges, indicating ferromagnetism as due to magnetic proximity effect. The estimated moment of $0.16 \pm 0.08$ $\mu_B$ leads to a clearly observable signal here, while no dichroic signal is evident for the pure monolayer without Fe coverage, putting stringent constraints on the magnitude of any possible magnetic moment in pristine VSe$_2$. Indeed, our results are thus in strong support of recent observations that the pristine MBE-deposited VSe$_2$ monolayer is not ferromagnetic.

For the proximity-coupled system studied here, the combination of XAS, PES, LEED, ARPES and angular dependent PES shows how the structural and chemical order of interfacial VSe$_2$ is endangered upon Fe deposition. A tendency of V to metallize towards a $V^0$ state, originating from the Se propensity to migrate towards the surface, is observed.

Our results show that the chemical stability of ML-VSe$_2$ upon deposition of a metallic ferromagnetic layer may be partially lost. At the meantime, the clear proximity-induced coupling at the interface between V and Fe motivates to further explore different ferromagnetic/2D-TMDC heterostructures.

## Acknowledgements

This work has been partially performed in the framework of the nanoscience foundry and fine analysis (NFFA-MIUR Italy Progetti Internazionali) project. We gratefully acknowledge support from The Leverhulme Trust (Grant No. RL-2016-006) and The Royal Society.